\documentclass[11pt,a4paper]{article}
\pdfoutput=1

\usepackage{bm}
\usepackage{amsmath,amssymb}
\usepackage{cite}

\usepackage{graphicx}
\usepackage{color}
\usepackage{upgreek}

\usepackage{hyperref} 
\hypersetup{colorlinks=true, citecolor=blue, filecolor=black, linkcolor=blue, urlcolor=blue, pdfpagemode=UseNone}

\numberwithin{figure}{section}
\numberwithin{equation}{section}

\newcommand{\be}{\begin{equation}}
\newcommand{\ee}{\end{equation}}
\newcommand{\bea}{\begin{eqnarray}}
\newcommand{\eea}{\end{eqnarray}}
\def\beal#1\eeal{\begin{align}#1\end{align}}   
\def\besp#1\eesp{\begin{multline}#1\end{multline}} 

\newcommand{\Tillde}[1]{#1}

\newcommand\ie{\textit{i.e.}\ }
\newcommand\eg{\textit{e.g.}\ }
\newcommand\cf{\textit{cf.}\ }

\newcommand{\aka}{{a.k.a.}\ }

\newcommand{\viz}{{\it viz.}\ }

\newcommand{\nn}{\nonumber}
\newcommand{\RR}{{\mathbb R}}
\newcommand{\dd}[2]{\delta_{\!\phantom{(} #1}^{\!(#2)}\!(\varphi)}

\begin{document}

\begin{titlepage}

\begin{center}
{\huge \bf Provable properties of asymptotic safety in $f(R)$ approximation}

\end{center}
\vskip1cm


\begin{center}
{\bf Alex Mitchell, Tim R. Morris and Dalius Stulga}
\end{center}

\begin{center}
{\it STAG Research Centre \& Department of Physics and Astronomy,\\  University of Southampton,
Highfield, Southampton, SO17 1BJ, U.K.}\\
\vspace*{0.3cm}
{\tt  A.Mitchell-Lister@soton.ac.uk, T.R.Morris@soton.ac.uk, D.Stulga@soton.ac.uk}
\end{center}

\begin{abstract}
We study an $f(R)$ approximation to asymptotic safety, using a family of non-adaptive cutoffs, kept general to test for universality. Matching solutions on the four-dimensional sphere and hyperboloid, we prove properties of any such global fixed point solution and its eigenoperators. For this family of cutoffs, the scaling dimension at large $n$ of the $n^\text{th}$ eigenoperator, is $\lambda_n\propto b\, n\ln n$. The coefficient $b$ is non-universal, a consequence of the single-metric approximation. The large  $R$ limit is universal on the hyperboloid, but not on the sphere where cutoff dependence results from certain zero modes. For right-sign conformal mode cutoff, the fixed points form at most a  discrete set. The eigenoperator spectrum is quantised. They are square integrable under the  Sturm-Liouville weight. For wrong sign cutoff, the fixed points form a continuum, and so do the eigenoperators unless we impose square-integrability. If we do this, we get a discrete tower of operators, infinitely many of which are relevant. These are $f(R)$ analogues of novel operators in the conformal sector which were used recently to furnish an alternative quantisation of gravity.
\end{abstract}

\vskip4cm

\end{titlepage}

\tableofcontents

\newpage

\section{Introduction}
\label{sec:Intro}

The problem of perturbative non-renormalizability of gravity has spawned many new approaches to quantising gravitational interactions, of which the hypothesis of asymptotic safety is one of the most conservative. It is based on the familiar framework of quantum field theory and does not require introduction of new fields or structures, see \eg \cite{Percacci:2017fkn,Reuter:2019byg}.
The idea, first introduced by Weinberg \cite{Hawking:1979ig}, relies on the existence of an interacting fixed point that controls the behaviour of gravity at high energies, resulting in a non-perturbatively renormalizable theory.

The main tool used to explore this possibility is a functional renormalization group equation,  pioneered by Wilson \cite{Wilson:1973}, who called it the Exact RG (Renormalization Group). The version most often used is for the effective average action $\Gamma_k$, which is the Legendre effective action  modified with an IR cut-off $\mathcal{R}_k$, or equivalently the Legendre transform of the Wilsonian effective action \cite{Morris:1993}. It satisfies \cite{Wetterich:1992,Morris:1993}:
\begin{equation}
\label{ERG}
    \partial_t\Gamma_k=\frac{1}{2}\text{STr}\Big[(\Gamma^{(2)}_k+\mathcal{R}_k)^{-1} \partial_t\mathcal{R}_k \Big]\,,
\end{equation}
where $t=\ln(k/\mu)$ is the so-called RG time, $\mu$ being the usual arbitrary physical energy scale, STr is the space-time trace taking into account statistics of anticommuting fields, and $\Gamma^{(2)}_k$ is the Hessian: the second functional derivative of $\Gamma_k$ with respect to the fields.

Solutions of this equation determine the flow of the infinite number of effective couplings that parametrise the most general effective action. (The space spanned by all these couplings is known as theory space.) 
Exact solutions would require solving an infinite number of coupled differential equations and therefore seem out of reach in any realistic setting, particularly so for quantum gravity. Nevertheless, since the pioneering work on the simplest (the Einstein-Hilbert) truncation \cite{Reuter:1996}, 
a lot of evidence has been gathered in support of asymptotic safety by model approximations that  in particular suitably truncate the theory space to a finite number of couplings,  as summarised in \eg the review ref. \cite{Bonanno:2020bil}. An important step beyond this is to include an infinite number of couplings. To date this has been realised through versions of $f(R)$ approximations where the effective Lagrangian is approximated to be a function of the scalar curvature $R$ \cite{Codello:2007bd,Machado:2007,Codello:2008,Benedetti:2012,Demmel:2012ub,Demmel:2013myx,Benedetti:2013jk,Demmel:2014hla,Demmel:2014fk,Demmel2015b,Ohta:2015efa,Ohta2016,Percacci:2016arh,Morris:2016spn,Falls:2016msz,Ohta:2017dsq}:
\be
\label{fR}
  \Gamma_k=\int\! d^dx\,\sqrt{g}\,f_k(R)
\ee
(plus ghost and auxiliary field terms), and also through some closely related approximations \cite{Percacci:2015wwa,Labus:2015ska,Eichhorn:2015bna,Alkofer:2018baq,Burger:2019upn,Kluth:2020bdv}. In fact, the high order finite dimensional truncations \cite{Falls_2014,Falls:2018ylp,Falls:2017lst,Kluth:2020bdv} were developed by taking examples of these $f(R)$ equations and then further approximating to polynomial truncations. The full $f(R)$ approximations are complicated partial differential equations so if no further approximation is applied, one is left to explore them numerically,\footnote{Some quadratic fixed point solutions can be found for special choices of parameters in the cutoff in \cite{Ohta:2015efa}.} supported by analysis in certain regions. One analytical approach is however particularly powerful, namely to solve the RG equations asymptotically at large curvature $R$ \cite{Dietz:2012ic,Gonzalez_Martin_2017}. Actually this technique is sufficient on its own to allow one to draw definitive conclusions about both the nature of the fixed points and their eigenoperator spectra in a given $f(R)$ approximation \cite{Dietz:2012ic,Gonzalez_Martin_2017}. It was adapted from studies of scalar field theories in derivative expansion approximation, where it had already proved to be powerful \cite{Morris:1994ki,Morris:1994ie,Morris:1994jc,Morris:1996nx,Morris:1997xj,Morris:1998}, and in ref. \cite{Dietz:2016gzg} it was also applied to the so-called conformal sector of quantum gravity. 

Another analytic approach that allows one to draw significant general conclusions is provided by Sturm-Liouville (SL) theory (see \eg \cite{Ince1956}). This was first demonstrated in scalar field theory \cite{Morris:1996xq}, while the tight theoretical structure that SL theory provides, lies behind the novel quantisation of gravity developed in refs. \cite{Morris:2018mhd,Kellett:2018loq,Morris:2018upm,Morris:2018axr,first,second}. In ref. \cite{Benedetti:2013jk}, SL analysis was used to prove properties of the eigenoperator spectrum in $f(R)$ approximations with \emph{non-adaptive} cutoff,\footnote{Contrast \emph{adaptive} cutoffs
which closely mimic the Hessian and thus also depend on $f_k(R)$, \cf sec. \ref{sec:fReqs} and \cite{Benedetti:2013jk}.}
namely that around any fixed point there are a finite number of relevant eigenoperators while the irrelevant eigenoperators form a tower whose scaling dimensions tend to infinity.

In this paper we combine both of these powerful analytical approaches to test the assumptions that are required for SL theory to apply $f(R)$ approximations, and to learn significantly more about the properties of asymptotically safe fixed points within these approximations. In particular we are able to test the extent to which results that should be universal are actually independent of the choice of cutoff, pointing to particular steps that need improving, and we are able to derive analytically the scaling dimensions of the eigenoperators at large dimension.

The structure of the paper is as follows. In the next section, following \cite{Benedetti:2013jk}, we set out the form of the flow equation, fixed point equation and eigenoperator equation. We discuss some of the choices to be made in particular for the endomorphism parameters $\alpha_s$, the choice of sign for the cutoff in the conformal factor sector, and the choice of background manifold. In sec. \ref{sec:sphere} we develop the equations in the case that the latter is a four-sphere, and explain further our choice of exponential cutoff for the common profile $r(z)$. As $R\to0$, the equations go over to a flat space limit. This is derived and discussed in sec. \ref{sec:flat}, and in particular its implications for SL theory where the $R=0$ boundary presents an obstruction. We see that the only sensible option is to continue  into the four-dimensional hyperboloid through a kind of smooth topology change, as discussed further also in the Conclusions, sec. \ref{sec:conclusions}. 

To apply SL theory we need that the eigenoperators are square integrable under the SL weight. The  question is whether this makes sense in quantum gravity. This leads us naturally into sec. \ref{sec:asymp} where we derive the asymptotic behaviour of solutions at large $R$. Separately this allows us to characterise the nature of the fixed points and eigenoperators.  First, in sec. \ref{sec:hyperboloid}, we explain the setup for the equations on the hyperboloid. In particular we furnish the full constraints on $\alpha_0$. In sec. \ref{sec:largeRfp} we derive the large $R$ asymptotic behaviour of a fixed point solution $f(R)$, first on the sphere and then on the hyperboloid. We see that the sphere solution differs from that assumed in  \cite{Benedetti:2013jk} and is in fact dominated by cutoff effects as $R\to\infty$. Computed exactly, it ought to be universal, as it is in fact for the hyperboloid. We show that the culprit is the course-graining of certain zero modes (modes with vanishing modified Laplacian) on the sphere. 

On both sphere and hyperboloid we see that the asymptotic solution contains only one parameter. Perturbations about this would provide the other parameter but such perturbations are invalid because they grow too fast. Here we find a beautiful connection to SL theory:  asymptotically they coincide with the inverse of the SL weight, which we derive in this regime on both manifolds. The fact that the asymptotic solutions for $f(R)$ contain only one parameter, allows us to draw an important conclusion: there are at most a discrete set of fixed point solutions. 

Sec. \ref{sec:largeReigen} presents analogous findings for the eigenoperators. The valid solutions are those that grow asymptotically as a power of $R$, the invalid solutions grow asymptotically like the inverse of the SL weight. Validity is decided by requiring their RG evolution to be multiplicative in the large $R$ limit. Left only with the power-law solutions, the equations are overconstrained leading to quantised values, $\lambda_n$, for their scaling dimensions. It is now immediate to see (in sec. \ref{sec:squareint}) that the valid eigenoperators  coincide with those that are square-integrable under the SL weight, justifying the use of SL analysis. 

Thus we have its standard result, stated in sec. \ref{sec:Liouville},  which in this context is that the scaling dimensions are real, that there are only a finite number of (marginally) relevant eigenoperators (such that $\lambda_n\le4$) and infinitely many irrelevant operators whose scaling dimensions $\lambda_n\to\infty$. By mapping to so-called Liouville normal form, the asymptotic analysis provides us with the large distance behaviour of the corresponding potential. From there by a standard application of WKB analysis, we get the analytical form for the scaling dimension $\lambda_n$ as a function of $n$, in the limit $n\to\infty$. This result should be universal. In fact it is independent of all but one of the parameters. We see that the remaining dependence is an artefact of the single-metric approximation.

In sec. \ref{sec:negative} we show that the situation changes dramatically if we choose the sign of the cutoff to be negative for the conformal factor sector. The SL weight now grows asymptotically, fixed points form a continuum, and the eigenoperator spectrum also becomes continuous. We relate this to earlier findings in $f(R)$ approximations with adaptive cutoff and in conformally reduced gravity. We show that we can impose square-integrability under the SL weight, in which case the valid operators are the ones that decay asymptotically like the inverse SL weight. We compute their asymptotic scaling dimensions, and we see that these operators are $f(R)$-analogues of the 
$\dd{k}n$ eigenoperators pursued in \cite{Morris:2018mhd,Kellett:2018loq,Morris:2018upm,Morris:2018axr,first,second} as an alternative quantisation of quantum gravity.

Finally, in sec. \ref{sec:conclusions} we bring these strands together, describe a search for numerical solutions, compare to $f(R)$ approximations with adaptive cutoff, and draw our conclusions. 

\section{The f(R) equations}
\label{sec:fReqs}

As explained in ref. \cite{Benedetti:2013jk}, a crucial choice is to take the cutoff profile to be $f(R)$ independent.  Although it does not allow the simplifications gained by combining the optimised cutoff function \cite{Litim:2001} with adaptive cutoff profiles (and thus used almost exclusively in all other studies), it has two advantages. Firstly, the flow equation is then second order in $R$-derivatives, rather than third order, which is crucial for the proofs. Secondly,  also crucial for the proofs (and one would assume also allowing more accurate modelling of the physics - see the further discussion in the conclusions, sec. \ref{sec:conclusions}) it ensures that the resulting ODEs for the fixed point solution and eigenoperators are free of fixed singularities. 
We use the same flow equation formulated in ref. \cite{Benedetti:2013jk}, and the same notation except that here we will work exclusively with quantities already scaled by the appropriate power of $k$ to make them dimensionless, thus avoiding the need to signify them with tildes. 

The flow equation takes the form of a non-linear partial differential equation for $f_k(R)$ \cite{Benedetti:2013jk}:
\be 
\label{flow}
 \partial_t f_k(R) + 2E(R) 
    =\frac1V \left(\mathcal{T}_2+\mathcal{T}_0^{\Bar{h}}+\mathcal{T}_1^{Jac}+\mathcal{T}_0^{Jac}\right) \,,
\ee
where $E(R)$ happens to be the equation of motion that would be deduced from the action \eqref{fR}:
\be 
E= 2f_k(R)-Rf'_k(R)\,.
\ee
Here, $V$ is the volume of space-time (scaled by $k^4$).  The space-time traces are given by: 
\beal
\label{T2}
\mathcal{T}_2 &=\text{Tr}\left[\frac{\frac{d}{dt} \mathcal{R}^T_k(\Delta_2+\alpha_2R)}{-f'_k(R)\Delta_2-E(R)/2 +2\mathcal{R}^T_k(\Delta_2+\alpha_2R)}\right]\,,\\
\label{Th}
 \mathcal{T}_0^{\Bar{h}} &=\text{Tr}\left[\frac{8\frac{d}{dt} \mathcal{R}^{\Bar{h}}_k(\Delta_0+\alpha_0R)}{9f''_k(R)\Delta_0^2+3f'_k(R)\Delta_0+E(R) +16\mathcal{R}^{\Bar{h}}_k(\Delta_0+\alpha_0R)}\right]\,,\\
 \label{T1J}
\mathcal{T}_1^{Jac} &=-\frac{1}{2}\text{Tr}\left[\frac{\frac{d}{dt}\mathcal{R}_k^V(\Delta_1+\alpha_1R)}{\Delta_1+\mathcal{R}^V_k(\Delta_1+\alpha_1R)}\right] 
\eeal
\be
\label{T0J}
 \mathcal{T}_0^{Jac} =\frac{1}{2}\text{Tr}\left[\frac{\frac{d}{dt}\mathcal{R}^{S_1}_k(\Delta_0+\alpha_0R)}{\Delta_0+R/3+\mathcal{R}^{S_1}_k(\Delta_0+\alpha_0R)}\right]-\text{Tr}\left[\frac{2\frac{d}{dt}\mathcal{R}^{S_2}_k(\Delta_0+\alpha_0R)}{(3\Delta_0+R)\Delta_0+4\mathcal{R}^{S_2}_k(\Delta_0+\alpha_0R)}\right]\,.
\ee
As explained in secs. \ref{sec:sphere} -- \ref{sec:hyperboloid}, they can be written as sums or integrals over the eigenvalues of the Laplacian operators.
The latter are modified to combinations appearing naturally in the space-time traces on a four-sphere \cite{Benedetti:2012}: 
\be
\Delta_s = -\nabla^2-\beta^S_s R\,,\qquad\text{where}\qquad \beta^S_0 =\tfrac13\,,\quad \beta^S_1 =\tfrac14\,,\quad \beta^S_2 =-\tfrac16\,.
\label{eq:19}
\ee
where a term proportional to $R$ has been added, for scalar, vector, and tensor modes respectively. 
The cutoff function $r(z)$ must be non-negative monotonic decreasing, and vanishing in the limit  $z\to+\infty$. 
For simplicity the same function is chosen for all field components so that, when scaled by the appropriate power of $k$, the cutoff profile takes the form
\begin{equation}
    \mathcal{R}^{\phi}_k=
    c_{\phi}\,r(\Tillde{\Delta}_s+\alpha_s\Tillde{R}) \,,\label{eq:15}
\end{equation}
where $c_\phi$ is a free parameter. Note that an additional correction  is incorporated, this time with coefficient $\alpha_s$.  These $\alpha_s$ are chosen to ensure that all modes are integrated out as $k\to0$, \ie such that all modes have positive $\Delta_s+\alpha_sR$.  Their value must be determined from knowledge of the spectrum on the appropriate background manifold(s), so we return to this issue later.  When written in terms of dimensionless quantities, as is done here, the total differential of the cutoff with respect to $t$, takes the form 
\be 
\label{dR}
\frac{d}{dt}\mathcal{R}^\phi_k(z) = c_\phi m_\phi\, r(z)-2c_\phi\, zr'(z)\,,
\ee
where $m_\phi$ is the mass-dimension of $\mathcal{R}^\phi_k$ (the same dimension as the Hessian it is regularising). 

In these equations, $\phi$ labels the field component. These  are metric fluctuation modes, namely the transverse traceless mode ($\phi=T$) and the gauge-invariant trace mode \aka the conformal factor field \cite{Gibbons:1978ac} ($\phi=\bar{h}$), and transverse vector and scalar modes from Jacobians of the field decomposition ($\phi=V,S_1,S_2$). The ghost and longitudinal modes do not appear since they cancel each other in Benedetti's scheme \cite{Benedetti:2012}. 

Actually, choosing $r(z)$ to be the same for all these modes is more than just a question of simplicity. The modes are all either part of the metric itself or directly related to it via the change of variables or via BRST transformations. Although BRST invariance of the quantum field is badly broken in the single metric approximation, it is reasonable to assume that the approximation would be poorer if we chose to regulate the parts in substantially different ways.

The $c_\phi$ determine the sign of the cutoff terms in the functional integral. If we require convergence of the integral we need $c_\phi>0$. We insist on this for $\phi=T,V,S_1,S_2$.
The situation is less clear however for the conformal factor. At the classical level $f(R)\sim -R$ is just the Einstein-Hilbert action, and in this case the conformal factor has a wrong-sign kinetic term (Hessian). One can see this from the denominator of the $\mathcal{T}^{\bar{h}}_0$ trace, \eqref{Th}, where the Hessian would reduce to $\sim -\Delta_0$  in this case. Therefore the trace is non-singular and the Functional RG is well-defined, only for $c_{\bar{h}}<0$ \cite{Reuter:1996,Dietz:2015owa,Dietz:2016gzg,Morris:2018mhd}. At the quantum level and depending on the value of $R$, the Hessian can be of either sign \cite{Lauscher:2002sq}. Classically the Hessian can also be of either sign if for example one includes a positive $R^2$ term. (This is the so-called Starobinsky term, a physically acceptable modification of Einstein's gravity. It corresponds to incorporating a ``scalaron''  \cite{Starobinsky:1980te} at the classical level.)

In the adaptive cutoff scheme the sign adapts so as to always be consistent with the Hessian. In the non-adaptive scheme that we need to use here, we have to make a choice, which will mean that the Functional RG is only applicable in the regime where this choice is consistent. As we will see this choice profoundly influences  RG properties. Where we need to decide we will choose $c_{\bar{h}}>0$, as in ref. \cite{Benedetti:2013jk,Lauscher:2002sq}, which means however that this version of the flow equation does not describe the regime corresponding to perturbative quantisation of the Einstein-Hilbert action. Then at the end of the paper, in sec. \ref{sec:negative}, we show what happens if we take $c_{\bar{h}}<0$ instead.


One small advantage of using a non-adaptive cutoff profile is that, since it does not itself depend on $f_k(R)$, the only occurrence of the RG time derivative acting on $f(R)$ is the one on the LHS of the flow equation \eqref{flow}. The fixed point equation for $f_k(R)=f(R)$ is then just given by dropping this term from the LHS, yielding a non-linear second order ordinary differential equation for $f(R)$:
\be
    2E(R)
    =\frac1V\left(\mathcal{T}_2+\mathcal{T}_0^{\Bar{h}}+\mathcal{T}_1^{Jac}+\mathcal{T}_0^{Jac}\right)\,. \label{fixed}
\ee
Linearising around such a fixed point solution, and separating variables,
\be
\label{linearised}
  f_k(R)=f(R)+\epsilon\, v(R)\,\text{e}^{-\theta t}\,,
\ee
(where $\epsilon$ is a small parameter) gives a linear second order ordinary differential eigenvalue equation:
\be
  -a_2(R)\, v''(R)+a_1(R)\, v'(R)+a_0(R)\, v(R)=\lambda\, v(R)\,, \label{eigenop}
\ee
where the eigenvalue $\lambda=4-\theta$ is the scaling dimension of the eigenoperator $v(R)$ and
\be
\label{a2}
  a_2(R) =\frac{144c_{\bar{h}}}{V}\text{Tr}\left[\frac{\Delta_0^2(2r(\Delta_0+\alpha_0R)-(\Delta_0+\alpha_0R)r'(\Delta_0+\alpha_0R))}{\left\{9f''(R)\Delta_0^2+3f'(R)\Delta_0+E(R)+16c_{\bar{h}}r(\Delta_0+\alpha_0R)\right\}^2}\right]
\ee
\beal
\label{a1}
    a_1(R) &= 2R-\frac{16c_{\bar{h}}}{V}\text{Tr}\left[\frac{(3\Delta_0-R)(2r(\Delta_0+\alpha_0R)-(\Delta_0+\alpha_0R)r'(\Delta_0+\alpha_0R))}{\left\{9f''(R)\Delta_0^2+3f'(R)\Delta_0+E(R)+16c_{\bar{h}}r(\Delta_0+\alpha_0R)\right\}^2}\right] \nonumber \\
    &\qquad+\frac{2c_T}{V}\text{Tr}\left[\frac{(R/2-\Delta_2)(2r(\Delta_2+\alpha_2R)-(\Delta_2+\alpha_2R)r'(\Delta_2+\alpha_2R))}{\left\{-f'(R)\Delta_2-E(R)/2+2c_Tr(\Delta_2+\alpha_2R)\right\}^2}\right]_{\vphantom{Gap}_{\vphantom{Gap}}}\\
\label{a0}
    a_0(R) &=\frac{32c_{\bar{h}}}{V}\text{Tr}\left[\frac{(2r(\Delta_0+\alpha_0R)-(\Delta_0+\alpha_0R)r'(\Delta_0+\alpha_0R))}{\left\{9f''(R)\Delta_0^2+3f'(R)\Delta_0+E(R)+16c_{\bar{h}}r(\Delta_0+\alpha_0R)\right\}^2}\right] \nonumber \\
    &\qquad+\frac{2c_T}{V}\text{Tr}\left[\frac{(2r(\Delta_2+\alpha_2R)-(\Delta_2+\alpha_2R)r'(\Delta_2+\alpha_2R))}{\left\{-f'(R)\Delta_2-E(R)/2+2c_Tr(\Delta_2+\alpha_2R)\right\}^2}\right]\,.
\eeal
Notice that the trace in $a_2(R)$ is positive thanks to the properties of $r(z)$. 
This is the reason for the sign in \eqref{eigenop}, since $a_2$ then has the same sign as $c_{\bar{h}}$ and in particular is positive for our choice $c_{\bar{h}}>0$.
The RG eigenvalue $\theta$ is the scaling dimension of the corresponding coupling. It has positive/zero/negative real part if the eigenoperator $v(R)$ is relevant/marginal/irrelevant. 

One of our main goals is to explore the applicability of  SL theory to the eigenoperator equation \eqref{eigenop} and when applicable, use it to prove properties of the eigenoperator spectrum \cite{Morris:1996xq,Benedetti:2013jk}. The derivation of the traces assumes that the background metric corresponds to a Euclidean-signature space of maximal symmetry. 
Globally, discrete choices are still possible, for example the real projective space $RP^4$ ($R>0$), torii ($R=0$), and analogous manifolds when $R<0$. However, we will see that if SL theory is to be applicable, then the only sensible choice is to incorporate a kind of smooth topology change between  $R>0$, $R=0$ and $R<0$ spaces. This is not possible unless we take maximal symmetry to apply also globally, as is standard practice in asymptotic safety approximations. Then $R>0$ corresponds to the four-sphere, $R=0$ to $\RR^4$, and $R<0$ to the four-dimensional hyperboloid. 

\begin{table}[tbh]
    \centering
    \begin{tabular}{c|c|c}
         Spin s& Eigenvalue $\lambda_{n,s}$ & Multiplicity $D_{n,s}$  \\
         \hline
         \hline
        0 & $\frac{n(n+3)-4}{12}R$&$\frac{(n+2)(n+1)(2n+3)}{6}$ \\
        \hline
        1 & $\frac{n(n+3)-4}{12}R$ & $\frac{n(n+3)(2n+3)}{2}$ \\
        \hline
        2 & $\frac{n(n+3)}{12}R$&$\frac{5(n+4)(n-1)(2n+3)}{6}$ \\
    \end{tabular}
    \caption{Multiplicities and eigenvalues for the four-sphere space-time traces. The sums for $\mathcal{T}_2$ and $\mathcal{T}_1^{Jac}$ begin at $n_\phi=2$, for $\mathcal{T}_0^{Jac}$ at $n_\phi=1$, and for $\mathcal{T}_0^{\Bar{h}}$ at $n_\phi=0$ \cite{Benedetti:2012}. }
    \label{tab1}
\end{table}

\subsection{Sphere}
\label{sec:sphere}

We start on the four-sphere, which was the space-time explicitly treated in \cite{Benedetti:2013jk}. It has space-time volume $V=384\pi^2/R^2$, and there the space-time traces are  sums over the discrete set of eigenvalues of the corresponding Laplacian:
\begin{equation}
    \text{Tr}\,W(\Delta_s)=\sum_{n=n_\phi}^\infty D_{n,s}\,W(\lambda_{n,s})\,. \label{eq:21}
\end{equation}
The multiplicities $D_{n,s}$, eigenvalues $\lambda_{n,s}$, and lowest index $n_\phi$,  are given in table \ref{tab1} (and its caption).

As is clear from the cutoff profile formula, \eqref{eq:15}, the $\alpha_s$ parameters allow us to shift the action of the cutoff up or down relative to the tower of eigenvalues,
so as to ensure all the modes are passed as $k$ is lowered to $k\to0^+$. 
Clearly this requires that the lowest mode $\lambda_{n_\phi,s}+\alpha_sR$ is  positive.
As noted in ref.  \cite{Benedetti:2013jk}, it is safe to choose $\alpha_2=0$ and $\alpha_1=0$, but to implement this condition in the physical scalar (\aka conformal factor) sector we need to choose\footnote{In this sense the modes are not treated equally. There appears to be no solution that does treat them `equally' at this level of detail, given constraints that we will also have to satisfy on the hyperboloid, \cf eqn. \eqref{betaH}.}  $\alpha_0>1/3$. 


At this point we recognise the need to specialise to \emph{smooth} (infinitely differentiable) cutoff functions $r(z)$. Given that the eigenvalues are discrete set, proportional to $R$, cutoff functions that are not smooth, for example the optimised one $r(z)=(1-z)\,\theta(1-z)$ \cite{Litim:2001}, will lead to points of limited differentiability which moreover accumulate as $R\to0$. It may still be possible to find a suitable \emph{weak} solutions to the fixed point and eigenoperator equations in this circumstance but, given that with a non-adaptive cutoff profile there is no advantage to using the optimised cutoff, there is no point in pursuing this possibility further. In fact one should bear in mind that cutoffs involving the Heaviside $\theta$ function have a number of related unpleasant effects\footnote{In real space the Kadanoff blocking functions are not truly quasi-local (they have power-law tails) and IR regulated vertices have no Taylor expansion in momentum (derivative expansion) beyond some low order.} that strictly speaking should rule them out as sensible choices, even if these problems are not obvious at current levels of approximation.  On the other hand, any smooth profile $r(z)$ will do if it decays sufficiently fast at large $z$. In our case we only need  to guarantee the convergence of the space-time traces above. Later we will specialise to the popular choice \cite{Wetterich:1992}
\be 
\label{Wetr}
r(z) = \frac{z}{\exp(az^b)-1}\,,\qquad a>0\,, b\ge1\,.
\ee
For any non-vanishing $R>0$ the sums are then rapidly convergent. However even if we restrict ourselves to four-spheres, we still need to understand the limiting case $R\to0^+$, which takes us to the boundary of this set. There, the sums go over to an integral and the equations go over to ones in flat space.

\subsection{Flat space}
\label{sec:flat}

This limit can be achieved by setting $p = n\sqrt{R/12}$, and then taking $R\to0$ whilst keeping $p$ fixed.
From table \ref{tab1}, it is clear that all the Laplacians $\Delta_{n,s}\to p^2$, \ie go over to their flat space limit where we recognise that $p$ is the flat space momentum. The multiplicities become $p^3(12/R)^{3/2}$ up to a numerical factor, while the cutoff profiles $\mathcal{R}^{\phi}_k\to c_\phi\, r(p^2)$. Putting all this together gives for the flow equation \eqref{flow},
\besp
\label{flatflow}
 \partial_t f_k(0)+4 f_k(0) = \frac1{8\pi^2}\int^\infty_0\!\!\! dp\, p^3 \Bigg\{  16c_{\bar{h}}\, \frac{2r(p^2)-p^2r'(p^2)}{9f''_k(0)\,p^4+3f'_k(0)\,p^2+2f_k(0)+16c_{\bar{h}}r(p^2)}\\
+10c_T\,\frac{r(p^2)-p^2r'(p^2)}{-f'_k(0)\,p^2-f_k(0)+2c_Tr(p^2)}
-3c_V \,\frac{r(p^2)-p^2r'(p^2)}{p^2+c_V r(p^2)}
\\
+c_{S_1} \frac{r(p^2)-p^2r'(p^2)}{p^2+c_{S_1}r(p^2)}
-4c_{S_2}\,\frac{2r(p^2)-p^2r'(p^2)}{3p^4+4c_{S_2} r(p^2)}\Bigg\}\,.
\eesp
For this to be well defined,
 $f_k(0)$, $f'_k(0)$, and $f''_k(0)$, need to be such that neither denominator vanishes (at some $p^2$) in its first two terms. This already provides strong constraints if the solution is to exist for all $k$, which however are soluble locally.
The strongest constraints arise if the cutoff function $r(z)$ diverges as $z\to0$, for example in the cases $b>1$ in \eqref{Wetr}. In the tensor mode trace, $c_T\,r(p^2)\to+\infty$ as $p\to0$ and thus the denominator is positive. On the other hand if $f'_k(0)$ is non-vanishing, as $p\to\infty$ the sign of the denominator is given by $-f'_k(0)$. Thus we see that to avoid a singularity we must always have $f'_k(0)\le0$ (strictly less than zero is the physically motivated choice since this corresponds to positive Newton's constant at zero momentum). Similarly we see that $f_k(0)$ is bounded above. From the $\bar{h}$ trace we see that since we chose $c_{\bar{h}}>0$, we must have $f''_k(0)>0$, 
while $f'_k(0)$ must also be bounded below by some negative value. 
If  $r(0)$ is finite, for example the case $b=1$ in \eqref{Wetr}, then other possibilities arise since $f'_k(0)$ can be positive if $f_k(0)>2c_T\,r(0)$, while the $\bar{h}$ trace would then only require that $f''_k(0)>0$. These considerations inform numerical searches, which we describe in sec. \ref{sec:conclusions}.

For the fixed point solution $f_k(R)=f(R)$, eqn. \eqref{flatflow} determines $f''(0)$ given boundary conditions $f(0)$ and $f'(0)$ such that all three lie within the bounds above. It then provides us with a Taylor expansion approximant to a putative fixed point solution:
\be 
\label{Taylor}
f(R) = f(0)+f'(0) R+\frac12 f''(0) R^2 +o(R^2)\,.
\ee
(In fact taking the expansion further is not straightforward since it then depends on the error in approximating the sums by integrals but these are not captured correctly by Euler-Maclaurin corrections.)
With these choices, the eigenoperator equation coefficients have finite limits:
\beal
\label{a2flat}
a_2(0) &= \frac{18c_{\bar{h}}}{\pi^2}\int^\infty_0\!\!\! dp\, p^7 \frac{2r(p^2)-p^2r'(p^2)}{\left\{9f''(0)\,p^4+3f'(0)\,p^2+2f(0)+16c_{\bar{h}}r(p^2)\right\}^2}\,,\\
\label{a1flat}
a_1(0) &= \int^\infty_0\!\!\! dp\, p^5 \Bigg\{ \frac{5c_T}{4\pi^2}\frac{r(p^2)-p^2r'(p^2)}{\left\{-f'(0)\,p^2-f(0)+2c_Tr(p^2)\right\}^2}\nn\\
&\qquad\qquad\qquad\qquad-\frac{6c_{\bar{h}}}{\pi^2}\frac{2r(p^2)-p^2r'(p^2)}{\left\{9f''(0)\,p^4+3f'(0)\,p^2+2f(0)+16c_{\bar{h}}r(p^2)\right\}^2}\Bigg\}\,,\\
a_0(0) &= \int^\infty_0\!\!\! dp\, p^3 \Bigg\{ \frac{5c_T}{4\pi^2}\frac{r(p^2)-p^2r'(p^2)}{\left\{-f'(0)\,p^2-f(0)+2c_Tr(p^2)\right\}^2}\nn\\
\label{a0flat}
&\qquad\qquad\qquad\qquad-\frac{4c_{\bar{h}}}{\pi^2}\frac{2r(p^2)-p^2r'(p^2)}{\left\{9f''(0)\,p^4+3f'(0)\,p^2+2f(0)+16c_{\bar{h}}r(p^2)\right\}^2}\Bigg\}\,,
\eeal
the eigenoperator equation itself then just being given by setting $R=0$ in \eqref{eigenop}. Furthermore $a_2(0)$ is non-vanishing since the integrand is positive definite.

The above implies that the SL weight function is finite and non-vanishing at $R=0$: 
\be 
\label{SLweight}
w(R)=\frac1{|a_2(R)|}\exp{-\int^R_0 \!\!\!\!dR'\, \frac{a_1(R')}{a_2(R')}}\,,
\ee 
(setting the lower limit in the integral to zero without loss of generality, and taking the modulus in the prefactor so that $\omega$ is positive whatever sign of $c_{\bar{h}}$ we choose.)
Multiplying the eigenoperator equation \eqref{eigenop} by the weight function (\aka the SL measure)
we can cast it in SL form:
\be
  -\left(a_2(R)w(R)v'(R)\right)'+w(R)a_0(R)v(R)=\lambda w(R) v(R)\,. \label{eq:26}
\ee
However SL properties only follow if the differential operator on the LHS is self-adjoint. Taking $v=v_j(R)$, multiplying by $v_i(R)$, and integrating over $R$, this means in particular that boundary terms must vanish when integrating by parts.  We see that we thus require the eigenfunctions to be square integrable under the weight function, and if we work only with fixed topology (here four-spheres), then for any two eigenfunctions $v_i(R)$ and $v_j(R)$, we 
get from the $R=0$ boundary:
\be 
\label{eigenzeroconds}
w(0) \left( v_i(0) v'_j(0) - v_j(0) v'_i(0)\right) =0\,,
\ee
the so-called bilinear concomitant.
Therefore we would have to choose all eigenoperators to satisfy a boundary condition, the most general form interpolating Dirichlet and Neumann: $\alpha v_i(0)+\beta v'_i(0)=0$ $\forall i$ (for some fixed $\alpha,\beta$). 
Such conditions lack any physical or other mathematical motivation, in particular in the full theory they cannot be respected beyond linearised order.
Our remaining option is to eliminate the $R=0$ boundary, requiring the solution to extend to all real values of $R$.

In fact we will get extra motivation for such choices when we analyse the number of fixed point solutions (or rather the dimension of the space of such solutions) in sec. \ref{sec:asymp}.
For this latter reason,  the $f(R)$ equations of ref. \cite{Benedetti:2012} were extended in ref. \cite{Dietz:2012ic} to all real $R$, by analytic continuation. Here we do not have the option of analytic continuation if we insist on using the same cutoff function $r(z)$ for all modes. The reason is that $\Delta_{n,s}$, being proportional to $R$, would change sign. Apart from the cutoff, this makes the denominator in $\mathcal{T}^{Jac}_1$, and in the $S_1$ term in $\mathcal{T}^{Jac}_0$, change sign, \cf  \eqref{T1J} and \eqref{T0J} respectively. By choosing sufficiently large $n$ in the modes in table \ref{tab1}, we see that the denominators will then vanish already at small negative $R$ making these traces ill-defined, unless we take $r(z)$ itself to be odd in $z$ (using \eg \eqref{Wetr} with $b=2$). However if we do take $r(z)$ odd, then instead the $S_2$ term in $\mathcal{T}^{Jac}_0$ will diverge already at small negative $R$ by similar arguments. (The other two traces also have their problems but since they involve $f(R)$, the demonstration is more involved.)

\subsection{Hyperboloid}
\label{sec:hyperboloid}

This leaves us with the remaining alternative, which is to match into the equations on a manifold with $R<0$. As explained earlier, we take for this the four-dimensional hyperboloid. Here $-\nabla^2$ is positive definite.  The volume $V$ is infinite, but the flow equation \eqref{flow} still makes sense since the space-time traces on the RHS trivially contain the same factor \cite{Camporesi:1994ga}:
\be 
    \frac1V\text{Tr}\,W(\Delta_s)= \frac{2s+1}{8\pi^2}\left(-\frac{R}{12}\right)^2\int^\infty_0\!\!\! d\lambda\left(\lambda^2+\left[s+\tfrac12\right]^2\right)\lambda\tanh(\pi\lambda)\, W(\Delta_{\lambda,s})\,;
 \ee
the spectrum is now continuous, indexed by $\lambda$:
\be 
\label{DeltaH}
\Delta_{\lambda,s} = \left(\lambda^2+s+\frac94\right)\left(-\frac{R}{12}\right) -\beta^S_s R
= -\frac{R}{12}\lambda^2-\beta^H_s R\,,
\ee
where thus
\be 
\label{betaH}
\beta^H_0 =\frac{25}{48}\,,\qquad \beta^H_1 =\frac{25}{48}\,,\qquad \beta^H_2 =\frac9{48}\,.
\ee
Recalling the reason for the extra endomorphism  in the cutoff profiles \eqref{eq:15}, we see that we can continue to set $\alpha_2=0$ and $\alpha_1=0$ as we wanted for the four-sphere, but the lower bound $\alpha_0>1/3$ is now joined by an upper bound
 $\alpha_0<25/48$ \cite{Benedetti:2013jk} so that all modes $\Delta_{\lambda,0}+\alpha_0 R > 0$.
 
The equations at the $R\to0^-$ boundary of this set of hyperboloids, are found by setting $p = \lambda\sqrt{-R/12}$ and holding $p$ fixed, so that once again the Laplacian goes over to its flat space expression $\Delta_{\lambda,s}\to p^2$. It is straightforward to verify that the flow equation \eqref{flow} and eigenoperator equation coefficients \eqref{a2}--\eqref{a0} then go over to the flat space expressions \eqref{flatflow} and \eqref{a2flat}--\eqref{a0flat} respectively. Thus we see that the flow, fixed point, and eigenoperator, equations can be smoothly defined over the combined set of all four-spheres, all four-hyperboloids, and $\RR^4$. The $R>0$ and $R<0$ parts of the solutions can be made to match as Taylor expansions around $R=0$ up to the second derivative,  but not beyond that. In fact the hyperboloid has a straightforward smooth limit\footnote{
Solutions can be straightforwardly developed to all-orders in the Taylor expansion around $R=0$, with coefficients given by finite integral expressions over $p$ similar to those in \eqref{flatflow}, \eqref{a2flat}--\eqref{a0flat}.}
whereas $f'''(0)$ on the sphere side depends also on corrections involved in converting the sums over eigenvalues into integrals.
In this way we have incorporated a smooth topology change mechanism through  these three spaces.

To apply SL theory, we are left only to establish acceptable behaviour at large $R$. In particular, as we have already seen, we need the eigenoperators $v(R)$ to be square integrable under the SL weight $w(R)$. Whilst this condition is natural for SL theory, and was assumed in ref. \cite{Benedetti:2013jk} for that reason,  the question is whether this makes sense in quantum gravity.

\section{Asymptotic behaviour of solutions at large R}
\label{sec:asymp}

We therefore turn now to the asymptotic behaviour of solutions at large $R$. This large field analysis also allows us to characterise a number of aspects of the solution space for both fixed points and their eigenoperator spectrum 
\cite{Morris:1994ki,Morris:1994ie,Morris:1994jc,Morris:1995he,Morris:1996nx,Dietz:2012ic,Dietz:2016gzg,Gonzalez_Martin_2017} and in particular allows us to answer the question above. We will see from sec. \ref{sec:negative} that the answer depends very much on the choice of sign for $c_{\bar{h}}$.

\subsection{Large R dependence of fixed points and how to count them}
\label{sec:largeRfp}

We start with the asymptotic behaviour of the fixed point solution $f(R)$. We need to know this in order to establish the large $R$ behaviour of the coefficients $a_i(R)$ in the eigenoperator equation \eqref{eigenop} which in turn will allow us to analyse the asymptotic behaviour of the eigenoperators.  However as we will see, it is important also for determining features of the fixed point solution space. 

Beginning with the sphere, and given a rapidly decaying cutoff profile $r(z)$, at first sight one can neglect the traces on the RHS of the fixed point equation \eqref{fixed} at large $R$. One would then conclude  
that $f(R) = AR^2$ plus rapidly decaying corrections \cite{Benedetti:2013jk}, for some undetermined coefficient $A$, this being the solution of just the LHS, $E(R)=0$. However this is not correct because terms in the traces whose denominator would vanish without a cutoff, yield a contribution on the RHS proportional to 
\be 
\label{rhscontribs}
\frac1{\mathcal{R}_k^\phi(z)}\frac{d}{dt} \mathcal{R}_k^\phi(z) = m_\phi -2 z\frac{d\ln r(z)}{dz}\,,
\ee 
where we used \eqref{dR}. There are three such terms, the $n=0$ and $n=1$ components from $\mathcal{T}^{\bar{h}}_0$ and the $n=1$ $S_2$ component of $\mathcal{T}^{Jac}_0$. The last two have $z=\alpha_0 R$ and actually cancel each other, but that still leaves the first contribution (with $z=[\alpha_0-\tfrac13]R$). Recalling the factor $1/V=R^2/384\pi^2$ on the RHS of the fixed point equation \eqref{fixed}, we see from \eqref{rhscontribs} that this contribution is not subleading to $f(R) = AR^2$, invalidating this ansatz. Actually this analysis shows that $f(R)$ grows faster than $R^2$. With this assumption the $n=0,1,$ $\mathcal{T}^{\bar{h}}_0$ terms now rapidly vanish, so that the only contribution that survives on the RHS at large $R$, is the $n=1$ $S_2$ component of $\mathcal{T}^{Jac}_0$. Keeping just this term it turns out one can solve the fixed point equation in closed form, thus obtaining the correct asymptotic behaviour for general cutoff function $r(z)$:
\be 
\label{asympGen}
f(R) = \frac{5R^2}{768\pi^2} \ln\frac{R^2}{r(\alpha_0R)}+AR^2+o(R^2)\qquad\text{as}\quad R\to+\infty\,,
\ee
where we used \eqref{rhscontribs} and noted that terms that grow slower than $R^2$ will be generated by iterating this asymptotic solution to higher orders. The $\ln r$ term actually dominates, \ie the large $R$ behaviour is dominated by cutoff-dependent effects. For example using the cutoff \eqref{Wetr}, \viz $r(z)= z/(\exp(az^b)-1)$ such that $a>0$, $b\ge1$, we find:
\be 
\label{asympFPS}
f(R) = \frac{5a\alpha_0^b}{768\pi^2}R^{2+b}+\frac{5}{768\pi^2}R^2\ln R+ AR^2+\frac{16c_{\bar{h}}}{5ab(1+b)\alpha_0^b}\left(\alpha_0-\frac13\right)\,\mathrm{e}^{-a\left(\alpha_0-\frac13\right)^b R^b}+\cdots\,,
\ee
where the ellipses stand for faster decaying terms. Here we adjusted $A$ to absorb a contribution to $R^2$, and then substituted the solution back into the fixed point equation to isolate the next leading correction. (This exponentially decaying correction comes from the $n=0$ term in the $\mathcal{T}^{\bar{h}}_0$ trace. All other corrections decay faster  provided that $\alpha_0 < \frac56+\alpha_1$. This is satisfied thanks to the restrictions imposed below \eqref{betaH}.)

Recalling that $R$ is the dimensionless version, \ie the physical curvature divided by $k^2$, we see that the large
$R$ limit may be viewed as holding the physical curvature fixed and integrating out all modes by sending $k\to0$. Therefore it ought to provide us with (an approximation to) the physical Legendre effective action, \ie the universal physical equation of state as a function of $R$ \cite{Gonzalez_Martin_2017}. The cutoff dependence however obstructs any attempt to extract physics from this limit. 
This problem is not seen in the Local Potential Approximation in scalar field theory, where an approximation to the equation of state can be successfully computed in this way \cite{Morris:1996xq}, and as we will see it is not a problem on the hyperboloid. Since the issue arises from the fact that the $n=1$ $S_2$ modes in the scalar Jacobian have vanishing eigenvalue, it suggests that further research should be done to understand if/how these modes can be better treated on a sphere.

Although the fixed point equation \eqref{fixed} is a second order ODE free of fixed singularities, the asymptotic solution we have found contains only the one free parameter: $A$. It is important to ask where the other parameter has gone. To find out, we linearise the fixed point equation around the asymptotic solution.  This just gives the eigenoperator equation \eqref{eigenop} for a marginal deformation, $\delta\!f(R)=\epsilon\, v(R)$ \ie such that $\theta=0$ or equivalently $\lambda=4$. As a linear second order ODE, it must have two linearly independent solutions. These can be found in the large $R$ limit.  Inspecting \eqref{a2} -- \eqref{a0}, we note that $a_1(R) =2R$ to leading order, while both $a_2(R)$ and $a_0(R)$ vanish asymptotically. We can therefore neglect $a_0$ and write
\be 
\label{perturbedFP}
4\, \delta\! f(R) - a_1(R)\,\delta\! f'(R) = -a_2(R)\, \delta \!f''(R) \,.
\ee
We know one solution to this already: $\delta\!f(R) = \delta\!A R^2 +\cdots$, where the RHS is only involved in supplying one of the subleading corrections. The other solution must thus be such that at leading order, $\delta\!f''(R)$ cannot be neglected. This tells us higher derivatives dominate over lower derivatives so we know that for the other solution $\delta\!f(R)$ can instead be neglected (to leading order). The equation is then exactly soluble since it can be rewritten as
\be 
\label{othersol}
\frac{d}{dR}\ln\delta\!f'(R) = \frac{a_1(R)}{a_2(R)} \qquad\implies\qquad \delta\!f(R) = B\int^R\!\!\!\!dR'\, \exp\! \int^{R'} \!\!\!\!dR''\, \frac{a_1(R'')}{a_2(R'')}\,,
\ee
where $B$ is the putative missing parameter.
For the explicit form we need $a_2$. It gets its leading contribution from the same source as the leading correction \eqref{asympFPS} to the terms displayed in \eqref{asympGen}. For the same cutoff choice \eqref{asympFPS}, we find asymptotically 
\be 
\label{asympa2S}
a_2(R) = \frac{24576\pi^2c_{\bar{h}}}{25ab(1+b)^2\alpha_0^{2b}}\left(\alpha_0-\frac13\right)^{1+b}\!\!\! R^{1-b} \,\mathrm{e}^{-a\left(\alpha_0-\frac13\right)^b R^b} +\cdots\,.
\ee
Recalling that $a_1 =2R$ to leading order, we can evaluate the integrals by successive integration by parts, as an asymptotic series and where each term is given in closed form. Since we will use this strategy many times let us sketch it on the indefinite integral:
\be 
\label{trick}
\int\!\!dR\, G(R) \,\mathrm{e}^{F(R)} = \frac{G(R)}{F'(R)}\,\mathrm{e}^{F(R)} - \int\!\! dR\, \left(\frac{G(R)}{F'(R)}\right)'\,\mathrm{e}^{F(R)}\,.
\ee
If $F(R)$ grows at least as fast as $R$ for large $R$, where $F$ is either sign, and $G(R)$ grows or decays slower than an exponential of $R$, then the integral on the right is subleading compared to the integral on the left. Iterating this identity then evaluates the integral in the large $R$ limit as $\mathrm{e}^{F(R)}$ times an asymptotic series, the first term on the RHS being the leading term. 

Using \eqref{asympa2S} this allows us to evaluate the inner integral in \eqref{othersol}. Its  exponential is then the integrand for the outer integral, such that the asymptotic series now provides subleading multiplicative corrections. Up to such corrections, the integrand is actually $1/\omega(R')$, as can be seen from eqn. \eqref{SLweight}. 
Applying the same integration by parts strategy to the outer integral does not change the leading exponential behaviour, and thus we see that up to subleading multiplicative corrections
$\delta\!f(R) \sim {B}/{\omega(R)}$
where we find the SL weight in the same approximation to be:
\be
\label{omegaS}
\omega(R) \sim 
\exp \left\{ -
\frac{25(1+b)^2\alpha_0^{2b}}{12288\pi^2c_{\bar{h}}}\left(\alpha_0-\frac13\right)^{-1-2b}\!\!\! R \,\mathrm{e}^{a\left(\alpha_0-\frac13\right)^b R^b}
\right\}\,.
\ee
 
Notice that the sign of $c_{\bar{h}}$ is crucial. Assuming $c_{\bar{h}}>0$, the linearised perturbation $\delta\!f(R) \sim {B}/{\omega(R)}$ is a rapidly growing exponential of an exponential. Taking the $R\to+\infty$ limit, it is not a small perturbation to our previous result \eqref{asympFPS}, no matter how small we choose $B$, thus invalidating the procedure used to derive it.\footnote{It can be understood as the linearised precursor to the solution ending in a (movable) singularity \cite{Morris:1994ki,Morris:1994ie,Morris:1994jc,Morris:1995he,Morris:1996nx,Dietz:2012ic,Dietz:2016gzg,Gonzalez_Martin_2017}.} Evidently it cannot itself satisfy the fixed point equation asymptotically (it would have to solve just the LHS to do that).
Therefore there is asymptotically only a one-parameter set of solutions namely \eqref{asympFPS}. 

The dimension of the fixed point solution space is determined by the asymptotic behaviour \cite{Gonzalez_Martin_2017}, thus unless further conditions are imposed we have (some discrete number of) lines of fixed points. 
Since it is not sustainable to try and impose a condition at $R=0$, \cf the discussion on eigenoperators below \eqref{eigenzeroconds},  we need to continue through smooth topology change (as defined at the end of sec. \ref{sec:fReqs}) into the hyperboloid side, if we are to reduce the dimension of the fixed point solution space from the current phenomenologically disappointing answer.

Turning to the hyperboloid, the situation is much more straightforward. The assumption that for rapidly decaying cutoff profile $r(z)$ one can neglect the traces \eqref{T2} -- \eqref{T0J} at large (negative) $R$, is now correct for the ansatz $f(R) = AR^2$,  thanks to the Laplacian eigenvalues \eqref{DeltaH} being bounded below sufficiently by the positive endomorphisms to avoid vanishing denominators, \cf \eqref{betaH}. Since the ansatz solves the LHS of the fixed point equation, it forms the start of the large $R$ asymptotic series solution. The traces provide corrections that decay thanks to the cutoff profiles' dependence on $\Delta_{\lambda,s}+\alpha_s R >  (\beta^H_s-\alpha_s) |R|$. From the $\alpha$ and $\beta$ parameter values, \cf \eqref{betaH} and below it, we see that 
\be
0<\beta^H_0-\alpha_0 < 9/48 = \beta^H_2 -\alpha_2 < \beta^H_1 -\alpha_1\,
\ee
and thus the leading corrections come from the scalar traces $\mathcal{T}^{\bar{h}}_0$ and $\mathcal{T}^{Jac}_0$. From the power of $\Delta_0$ in \eqref{T0J} it is the $S_1$ part that is leading. After some tedious manipulation we find:\footnote{Recall that $\alpha_0<25/48$.}
\be 
\label{asympFPH}
f(R) = AR^2 +\frac{c_{S1}}{96\sqrt{3\pi a^3b^3}}\left(\frac{25}{48}-\alpha_0\right)^{\frac{5-3b}{2}}\!\!\!\left(-R\right)^{2-\frac{3b}{2}}\left\{1+ O\left(|R|^{-\frac12}\right)\right\}\,\mathrm{e}^{-a\left[\left(\alpha_0-\frac{25}{48}\right)R\right]^b}+\cdots\,,
\ee
as $R\to-\infty$, all scalar traces (thus also $A$) contributing to the $O\left(|R|^{-\frac12}\right)$ term, and the ellipses standing for terms with faster decaying exponentials. 
Again we ask where the other parameter has gone. The analysis proceeds in a similar fashion to that on the sphere. We have again the asymptotic perturbed fixed point equation \eqref{perturbedFP} except now:
\be 
\label{asympa2H}
a_2(R)= \frac{4c_{\bar{h}}}{81A^2\sqrt{3\pi ab}}\left(\frac{25}{48}-\alpha_0\right)^{\frac{5-b}{2}}\!\!\!\left(-R\right)^{1-\frac{b}{2}}\,\mathrm{e}^{-a\left[\left(\alpha_0-\frac{25}{48}\right)R\right]^b}+\cdots
\ee
(the ellipses being faster decaying terms). A small perturbation to \eqref{asympFPH} gives 
 \eqref{othersol}, and thus we have $\delta\!f(R) \sim B/\omega(R)$ again, except now the SL weight is
\be 
\label{omegaH}
\omega(R) \sim \exp\left\{ -\frac{81A^2}{2c_{\bar{h}}} \sqrt{\frac{3\pi}{ab}}\left(\frac{25}{48}-\alpha_0\right)^{-\frac{b+5}{2}}\left(-R\right)^{1-\frac{b}{2}}\,\mathrm{e}^{a\left[\left(\alpha_0-\frac{25}{48}\right)R\right]^b}
\right\}
\ee
(where again we neglect also subleading multiplicative terms).
As $R\to-\infty$, such a $\delta\!f(R)$ is a rapidly growing exponential of an exponential, and thus asymptotically we have only the one-parameter set of solutions \eqref{asympFPH}.


These results allow us to draw an important conclusion. Each of the hyperboloid and sphere asymptotic solutions impose one constraint.\footnote{\textit{E.g.} $Rf'(R)-2f(R)=Rf'_{asy}(R)-2f_{asy}(R)$, for some suitably large $R$, where $f_{asy}$ is \eqref{asympFPS} for $R>0$, or \eqref{asympFPH} for $R<0$, and the RHS has no free parameters since the $AR^2$ term is cancelled out in this linear combination.}  Since we thus have two boundary conditions imposed on a second order ordinary differential equation we have at most a discrete set of solutions. \textit{A priori} this could be no fixed point, or a unique fixed point (the phenomenologically preferred answer), a larger number of fixed points, or a countable infinity of fixed points. As we will see in sec. \ref{sec:negative}, the conclusion is very different if we choose the conformal factor cutoff to be negative, \ie $c_{\bar{h}}<0$.

\subsection{Large R dependence of eigenoperators}
\label{sec:largeReigen}

Since the eigenoperator equation \eqref{eigenop} is linear and second order, there are guaranteed to be two independent solutions for any RG eigenvalue $\lambda$. Whether they are acceptable or not, crucially depends on their large field behaviour \cite{Morris:1996nx,Morris:1996xq,Morris:1998,Bridle:2016nsu,Dietz:2016gzg}: in particular whether for small but fixed $\epsilon$ the exponential dependence in RG time in \eqref{linearised} remains valid at large $R$. In scalar field theories this criterion explains why the correct eigenoperator solutions are the ones with power-law large field behaviour and thus why the RG eigenvalues are quantised \cite{Morris:1994ki,Morris:1994ie,Morris:1994jc,Morris:1995he,Morris:1996nx,Morris:1996xq,Morris:1998,Bridle:2016nsu}. It was also applied in ref. \cite{Dietz:2016gzg} to determine the eigenoperator spectrum around non-trivial fixed points in a conformal truncation to quantum gravity, and in ref. \cite{Dietz:2012ic} to an $f(R)$ approximation with adaptive cutoff  \cite{Benedetti:2012}.

We have just derived the asymptotic behaviour of $f(R)$ for a fixed point solution to the flow equation \eqref{flow} with non-adaptive cutoff.  Substituting this into the corresponding  eigenoperator equation \eqref{eigenop} allows us to determine the large $R$ behaviour of solutions $v(R)$. We will use the above insight to determine which of these solutions are valid. 

In fact, since with non-adaptive cutoff, RG time derivatives of $f_k(R)$ appear only the once, as $\partial_tf_k(R)$ on the LHS of the flow equation \eqref{flow}, one can immediately read off from the asymptotic form of the perturbed fixed point equation  \eqref{perturbedFP}, the corresponding asymptotic form of the eigenoperator equation \eqref{eigenop}:
\be 
\label{asympeigenS}
\lambda\, v(R) - 2R\,v'(R)  = -a_2(R)\,v''(R)\,,
\ee
where asymptotically $a_2$ is given by \eqref{asympa2S} or \eqref{asympa2H} as appropriate.
One solution solves just the LHS:
\be 
\label{eigenpow}
v(R) \propto |R|^{\frac\lambda2}+\cdots\,,
\ee
where the ellipses stand for subleading corrections including those supplied by the RHS, and we note that the solution is determined only up to a constant of proportionality. The other solution must be such that at leading order, $v''(R)$ cannot be neglected. For the same reasons as before, asymptotically the ODE then collapses to \eqref{othersol} (with $\delta\!f$ replaced by $v$)
and thus these solutions satisfy $v(R)\sim1/\omega(R)$,
with $\omega(R)$ being given by \eqref{omegaS} and \eqref{omegaH} on the sphere and hyperboloid respectively.

Now we ask whether these solutions are actually valid. 
The linearised solution  \eqref{linearised} is meant to describe the RG flow `close' to the fixed point. For any fixed $\epsilon$, if $|v(R)/f(R)|\to\infty$ as $R\to\pm\infty$ that is not necessarily true since linearisation is no longer valid. In this case we set
\be 
\label{pertgen}
f_k(R)=f(R)+\epsilon\,v_k(R)\,,
\ee 
and ask for the correct evolution for $v_k(R)$ at large $R$. We see that for large negative $R$ we can neglect the RHS of the flow equation \eqref{flow}. For large positive $R$ we can neglect the RHS of the flow equation except  for the $n=1$ $S_2$ component of $\mathcal{T}^{Jac}_0$, which however just cancels the contributions from the LHS that grow faster than $R^2$ resulting from $f(R)$, \cf \eqref{asympFPS}. Since in fact the $O(R^2)$ part of $f(R)$ also vanishes from the LHS  (on both sphere and hyperboloid), we see that  in the large $R$ regime we have 
\be 
\partial_t v_k(R)-2R\,v'_k(R)+4\,v_k(R) = o( R^2)\,.
\ee
Any part of $v_k(R)$ growing at least as fast as $R^2$ is then easily solved for, and gives mean-field evolution involving some arbitrary function $v$:
\be 
\label{tgen}
v_k(R) = {\rm e}^{-4t}\, v(R\,{\rm e}^{\,2t})+ o( R^2)\,.
\ee
It will be the same function $v$ that we introduced in the linearised solution \eqref{linearised} if we require as boundary condition, $v_k(R)=v(R)$ at $k=\mu$. The question that remains is whether the RG evolution \eqref{tgen} is consistent with what we were assuming by linearising.

For the power-law solution \eqref{eigenpow}, linearisation is valid at large $|R|$ if and only if $\lambda\le4$. This follows from the hyperboloid fixed point asymptotics \eqref{asympFPH},  the sphere side \eqref{asympFPS} requiring only the weaker constraint, $\lambda\le4+2b$.  On the other hand if $\lambda>4$, 
we use the general perturbation \eqref{pertgen}, finding the solution \eqref{tgen}. Substituting the explicit form \eqref{eigenpow} of the boundary condition we get
\be 
\label{backtolinearised}
v_k(R) = v(R)\,\text{e}^{-\theta t} + o( R^2)\,,
\ee
where $\theta=4-\lambda$, \ie we reproduce the linearised solution \eqref{linearised}. We conclude that asymptotically, power-law eigenoperators \eqref{eigenpow} are valid solutions for any $\lambda$. Their $t$ evolution is multiplicative and given by the flow of a conjugate coupling $g(t)=\epsilon\,\text{e}^{-\theta t}$, \cf \eqref{linearised}.

On the other hand, the solutions that behave asymptotically as $v(R)\sim1/\omega(R)$, are growing exponentials of exponentials. Linearisation is not valid at large $|R|$, where the $t$ dependence is given instead by \eqref{tgen}. Now we cannot separate out the $t$ dependence. Therefore such perturbations cannot be regarded as eigenoperators evolving multiplicatively.
Excluding them leads to quantisation of the spectrum. The large $R$ dependence \eqref{eigenpow} provides a boundary condition on both the sphere and the hyperboloid side, and  linearity provides a further boundary condition since we can choose a normalisation \eg $v(0)=1$. These three conditions over-constrain the eigenoperator equation \eqref{eigenop} leading to quantisation of $\lambda$, \ie to a discrete eigenoperator spectrum. 

Again we will see in sec. \ref{sec:negative}, that the conclusion is very different if we choose the conformal factor cutoff to be negative, \ie $c_{\bar{h}}<0$.

\subsection{Square integrability under the Sturm-Liouville weight}
\label{sec:squareint}

Now we can return to the question posed at the end of sec. \ref{sec:fReqs}: whether it makes sense for eigenoperators $v(R)$ to be square-integrable under the Sturm-Liouville weight $w(R)$, \cf \eqref{SLweight}, which is the remaining condition that must be satisfied in order for SL theory to be applicable. 
We have seen that on both manifolds, $\omega(R)$ is rapidly decaying for large curvature. We saw that the eigenoperator solutions that are actually allowed are the ones that grow as a power, \eqref{eigenpow}. Now we see that they are square integrable under this measure. On the other hand the solutions $v(R)\sim1/\omega(R)$ that we already excluded on physical grounds, satisfy  $\omega(R)\, v^2(R)\sim 1/\omega(R)$ which thus diverges at large $R$. These perturbations are therefore not square integrable under the measure. We conclude that the condition of square-integrability picks out the correct solutions from the eigenoperator equation and that SL theory is therefore applicable. 

Although these formulae have been derived for the specific choice of exponential cutoff \eqref{Wetr}, it is immediate to see that these qualitative properties hold true for a wide range of  cutoffs, independent of their details. Indeed the fact that $a_2(R)$ is decaying for large $|R|$ with sign given by $c_{\bar{h}}$, and that $a_1(R)=2R$ plus decaying terms, is enough to ensure that $\omega(R)$ for $c_{\bar{h}}>0$ is a rapidly decaying exponential, as follows from its formula \eqref{SLweight}. This behaviour also ensures that $\delta\!f(R)$, the non-power-law solutions $v(R)$, and $1/\omega(R)$, are all equal up to subleading multiplicative corrections. 
In sec. \ref{sec:negative}, we will see that if  we choose $c_{\bar{h}}<0$, these solutions still hold but lead to profoundly different scenarios.

\section{Liouville normal form}
\label{sec:Liouville}

We have seen that SL theory is (only) applicable to the quantised spectrum of eigenoperators that have power-law asymptotic behaviour in $R$, given by \eqref{eigenpow}, and which we determined already from RG properties were the physical eigenoperators. The consequences of SL theory for this spectrum can be seen by a standard transformation that takes the linear second order ODE \eqref{eigenop} to so-called Liouville normal  form. For this case we set the coordinate to be (taking $x=0$ at $R=0$ without loss of generality):
\be
\label{x}
  x=\int_{0}^R\frac{1}{\sqrt{a_2(R')}}dR'\,.
\ee
It is well defined since we have seen that $a_2(R)$ is strictly positive at all finite $R$.  Furthermore since $a_2(R)$ vanishes at large $|R|$ we see that $x\to\pm\infty$ as $R\to\pm\infty$. 
Then defining the `wave-function'
\be
\label{psi}
  \psi(x)=a_2^{\frac{1}{4}}(R)\,w^{\frac{1}{2}}(R)\,v(R)\,,
\ee
\eqref{eigenop} becomes
\be
  -\frac{d^2\psi(x)}{dx^2}+U(x)\,\psi(x)=\lambda\, \psi(x)\,, \label{Schro}
\ee
which is nothing but the time-independent Schr\"odinger equation at energy $\lambda$ (and mass $\tfrac12$). This is Liouville normal form. After some manipulation, one finds that 
the potential is given by \cite{Benedetti:2013jk}:
\be
\label{U}
  U(x)=a_0+\frac{a_1^2}{4a_2}-\frac{a'_1}{2}+a'_2\Big(\frac{a_1}{2a_2}+\frac{3a'_2}{16a_2}\Big)-\frac{a''_2}{4}
\ee
(the terms on the RHS being functions of $R$).

In ref. \cite{Benedetti:2013jk}, it was noted that this potential has no singularities at finite $x$ whilst from the asymptotic behaviour of the $a_i(R)$, the second term dominates for $x\to\pm\infty$ such that $U(x)\to+\infty$, leading to the conclusion that there is only a quantised bound-state energy spectrum $\lambda=\lambda_n$ ($n=0,1,2,\cdots$) bounded from below with the only accumulation point at infinity (following standard analysis of its Green's function, see \eg \cite{BerezinShubin}). In other words there are only a finite number of (marginally) relevant couplings such that $\theta_n=4-\lambda_n\ge0$, and infinitely many irrelevant couplings. These latter have scaling dimensions $\theta_n\to-\infty$ as $n\to\infty$.

There is a hidden assumption here, namely that $\psi(x)$ has appropriate behaviour as $x\to\pm\infty$ for the Schr\"odinger equation interpretation to make sense. For this, $\psi(x)$ should be either square-integrable, corresponding to a bound state, or correspond to an unbound state such that $\psi(x)=\psi_k(x)\sim \text{e}^{ikx}$ as $x\to\pm\infty$ for some wave-number $k$. These latter are $\delta$-function normalisable, \ie can be chosen to satisfy $\int_x \psi_k(x)\psi_{k'}(x) = \delta(k-k')$. For this potential these latter solutions do not exist. As we have seen, there are other solutions however, but the missing solutions (which we have rejected on RG grounds) behave asymptotically as $v(R)\sim1/\omega(R)$. From \eqref{psi}, they grow rapidly as  $x\to\pm\infty$ (in fact exponentially) so are neither square-integrable nor $\delta$-function normalisable. 

On both sphere \eqref{asympa2S} and hyperboloid \eqref{asympa2H}, we can write
\be 
\label{aFG}
a_2(R) = \frac1{G^2(R)}\,\mathrm{e}^{-2F(R)}\,,
\ee
where $F$ and $G$ have the behaviour required for the identity \eqref{trick}. Thus we get asymptotically 
\be 
\label{asympx}
x = \frac{G(R)}{F'(R)}\,\mathrm{e}^{F(R)} +\cdots\,.
\ee
From \eqref{U} to leading order, we therefore have
\be 
\label{asympU}
U(x) = \frac{a_1^2}{4a_2} = \frac{R^2}{a_2(R)} = \left[RF'(R)\right]^2 x^2\,.
\ee
But from \eqref{asympa2S} and \eqref{asympa2H} we see that $R F'(R) = b\, F(R)$. Taking logs of \eqref{asympx}, we thus find
\be 
\label{largeU}
U(x) =\left( b\, x \ln |x| \right)^2 \left\{1+O\left(\frac{\ln\ln|x|}{\ln|x|}\right)\right\}\qquad\text{as}\qquad x\to\pm\infty\,.
\ee
It is interesting that the leading large $x$ behaviour of the potential is symmetric about the origin $x=0$, even though $U(x)$ is surely not. In particular the fact that at large $R$, $f(R)$ is universal on the hyperboloid but dominated by cutoff effects on the sphere, does not result in different behaviour in the corresponding large $x$ regime of the potential $U(x)$.
It is also interesting that this leading behaviour is close to being universal, in that the only cutoff dependence is through the parameter $b$, the power entering the exponential fall-off form in the cutoff \eqref{Wetr}. Unfortunately this still amounts to strong dependence. 
Actually this remaining dependence is an artefact of the single-metric approximation \cite{Reuter:1996,Bridle:2013sra}, one consequence of which is to conflate the background curvature dependence in the cutoff, in particular in $F$, with that of the quantum field.\footnote{In reality $\Gamma_k$ is a functional of both the background metric $g^B_{\mu\nu}$ and the quantum metric $g^Q_{\mu\nu}$. It is $g^Q_{\mu\nu}$ differentials that appear in the fixed point and eigenoperator equations, and thus it is also the behaviour at large $g^Q_{\mu\nu}$ that we are interested in. In a non-adaptive scheme as employed here, cutoff profiles such as \eqref{eq:15} should in reality not depend on $g^Q_{\mu\nu}$ but only on its field differentials, since the cutoffs  are meant to regularise Laplacians for these modes.}

From \eqref{largeU} we can find for the quantised spectrum the asymptotic behaviour of their  scaling dimensions at large $n$:
\be 
\label{largen}
\theta_n = -b\,(n\ln n)\left\{1+O\left(\frac{\ln\ln n}{\ln n}\right)\right\}\qquad\text{as}\qquad n\to\infty\,.
\ee
This follows by noting that large values of $\lambda_n$ closely obey the WKB formula for the Schr\"odinger equation \eqref{Schro}:
\be 
\label{wkb}
\int^{x_n}_{-x_n}\!\!\!dx\,\sqrt{\lambda_n-U(x)} = (n+\tfrac12)\pi\,.
\ee
The boundaries of the integral should be the classical turning points, \ie  the solutions to $\lambda_n=U(x)$.
However up to multiplicative corrections of order $\ln\ln x_n/\ln x_n$ these can be taken to be $\pm x_n$ where at the same level of approximation,
\be
\label{lambdan}
\lambda_n = U(\pm x_n) = \left( b \,x_n \ln x_n\right)^2\,.
\ee 
Substituting this and $x=x_n y$ into \eqref{wkb} gives 
\be 
\label{largexn}
I\,b\, x_n^2 \ln x_n^2 = \pi(2n+1)\,,
\ee
where the integral
\be 
I = \int^1_{-1}\!\!\!dy\,\sqrt{1-\frac{U(yx_n)}{\left( b x_n \ln x_n\right)^2}}\ =\
\int^1_{-1}\!\!\!dy\,\sqrt{1-y^2}\ =\ \frac\pi2\,,
\ee
again up to corrections of order $\ln\ln x_n/\ln x_n$.
Thus in the large $n$ limit, we can solve \eqref{largexn} in terms of the Lambert $W$ function, as $\ln x^2_n = W(4n/b)$  (using the fact that $W$ satisfies $W(z)\exp W(z)=z$).  Substituting this solution into \eqref{lambdan}, using the asymptotic expansion of $W(4n/b)$, and again neglecting multiplicative corrections of order $\ln\ln x_n/\ln x_n$ or smaller, gives \eqref{largen}.


Using polynomial truncations taken to very high order, the $\theta_n$ were accurately estimated up to $n=70$ in ref. \cite{Falls:2018ylp}, and found to closely fit $\theta_n \approx 2.91 - 2.042 n$. However these were computed in an adaptive cutoff version of the $f(R)$ approximation and using optimised cutoff \cite{Litim:2001}. Given the above strong dependence on cutoff profile we cannot make a sensible comparison, although we note that given the weak dependence of $\ln n$, and ignorance of the neglected corrections, $b\approx1$ would provide a reasonable match.

\section{Wrong sign cutoff in the conformal sector}
\label{sec:negative}

In this section we show what changes if we choose a negative cutoff for the conformal mode, \ie $c_{\bar{h}}<0$.


In sec. \ref{sec:flat} we analysed the constraints on $f_k(0)$, $f'_k(0)$, and $f''_k(0)$. For completeness we show how these change with $c_{\bar{h}}<0$. Recall that the strongest constraints arise if the cutoff function $r(z)$ diverges as $z\to0$. Then we showed that from the tensor mode trace we must have $f'_k(0)\le0$. Now that $c_{\bar{h}}<0$, from the $\bar{h}$ trace we need $f''_k(0)\le0$ to avoid a singularity. The equations are then consistent provided $f_k(0)$ is less than some positive bound. If $r(0)$ is finite then other possibilities again arise for example $f''_k(0)>0$ is possible provided $f_k(0)$ is sufficiently positive.

Much more interesting is the effect of negative cutoff on the space of fixed points and eigenoperators. Recall that the asymptotic behaviour of the fixed point solutions $f(R)$ is given in the first instance by  asymptotic series whose leading term is a power of $R$: \eqref{asympFPS} in the case of the sphere and \eqref{asympFPH} for the hyperboloid. These contain one parameter $A$ (a different value in general on the sphere or hyperboloid). However to determine the true number of parameters in the asymptotic solution, we study the linear perturbation $\delta\!f(R)$ to these asymptotic series, and find $\delta\!f(R)\sim B/\omega(R)$, where $\omega$ is the SL weight and is given by \eqref{omegaS} or \eqref{omegaH} on the sphere or hyperboloid respectively. The derivation is still correct if $c_{\bar{h}}<0$, but the SL weight is now a rapidly growing exponential of an exponential (on both sides). Thus the perturbation $\delta\!f(R)\sim B/\omega(R)$
is a rapidly decaying exponential of an exponential. Whatever value of $B$ we choose, asymptotically our assumption that $\delta\!f(R)$ is much smaller than the series solutions, becomes ever more justified. Therefore asymptotically there is now a full two-parameter set of solutions, being to leading order precisely \eqref{asympFPS} or \eqref{asympFPH} as appropriate, plus  $B/\omega(R)$. Now these solutions impose no boundary conditions since at some appropriate large $R$, $f(R)$ and $f'(R)$ merely fix the values of the two parameters $A$ and $B$ in the appropriate asymptotic solution. Therefore if we have solutions they will be continuous `planes' of fixed points: two-dimensional sets parametrised by two real free parameters \cite{Gonzalez_Martin_2017}. 

It had already been noticed in $f(R)$ truncations with adaptive cutoff, that fluctuations from  the conformal factor govern the structure of the solutions  \cite{Dietz:2012ic,Demmel2015b}.
We now see that the reason is that it is intimately tied to the way this sector is regularised. For non-adaptive cutoff the choice $c_{\bar{h}}<0$ is the only one available for the Einstein-Hilbert truncation \cite{Reuter:1996} and for perturbative solution of the flow equation starting from the classical Einstein-Hilbert action (with or without a cosmological constant), but such a wrong-sign kinetic term plus wrong sign cutoff, leads to a continuum of fixed point solutions  \cite{Dietz:2016gzg}. The cause is the same as was found in these earlier papers, namely the fact that the fixed point large field asymptotic behaviour has the full set of parameters and thus imposes no boundary conditions. This effect is also seen \cite{Dietz:2012ic} in one formulation of $f(R)$ with an adaptive cutoff \cite{Benedetti:2012} and in some of the asymptotic solutions \cite{Gonzalez_Martin_2017} found in another formulation involving more fixed singularities \cite{Demmel2015b}.
It was already suggested in ref. \cite{Dietz:2012ic} that such a continuum of solutions is a reflection of the conformal mode instability. 

Similar conclusions are drawn for the eigenoperator spectrum around any such fixed point.
The exponential of exponential solutions $v(R)\sim1/\omega(R)$, to the eigenoperator equation \eqref{eigenop}, are now exponentially small at large $|R|$ and thus linearisation remains valid. Therefore we now have a continuous spectrum, with degeneracy two, for every value of $\lambda$.  Again, this effect has been seen before, in the same situations where a continuum of fixed points are found: in a conformal truncation \cite{Dietz:2016gzg}, and in $f(R)$ approximation with adaptive cutoff \cite{Dietz:2012ic}. 

Note that such a continuous spectrum of eigenoperators is consistent with there being a two-dimensional continuum of fixed points. Indeed the two eigenoperators with $\lambda=4$ are the exactly marginal operators $v(R)=\delta\!f(R)\sim R^2$ and $v(R)=\delta\!f(R)\sim 1/\omega(R)$ (for given sign of $R$) that move the system infinitesimally from one fixed point to another in this two-dimensional continuum.

%

The general eigenoperator with scaling dimension $\lambda$ grows as  $|R|^{\frac\lambda2}$ at large $|R|$, \cf \eqref{eigenpow}. They are thus not square-integrable under the SL weight. Although they have conjugate couplings that evolve multiplicatively at the linearised level, and are in this sense physical, we can choose to impose square-integrability as an extra condition. If we do so we exclude the power-law solutions.
This amounts to an extra \emph{quantisation condition} that is natural within the Wilsonian RG framework \cite{Morris:2018mhd}. Indeed without it the Wilsonian RG breaks down because there would be no sense in which an arbitrary linearised perturbation can be broken down uniquely into a convergent series expansion over operators of definite scaling dimension \cite{Dietz:2016gzg,Morris:2018mhd,Morris:toappear}.  The remaining solutions $v(R)\sim1/\omega(R)$ are exponentially decaying for both $R\to+\infty$ and $R\to-\infty$. Since for these, $\omega(R)\, v^2(R)\sim 1/\omega(R)$, they are square-integrable under the SL weight, and thus form a quantised spectrum. 

Their relation to the continuum of fixed points is novel in that it is no longer possible to move to any nearby fixed point by `switching on' marginal directions. Indeed we have at most one marginal operator now. Generically we will have none. 

The $a_2(R)$ coefficient \eqref{a2} changes sign under $c_{\bar{h}}\mapsto-c_{\bar{h}}$, but it
still decays exponentially at large $|R|$, as we see from \eqref{asympa2S} and \eqref{asympa2H} for sphere and hyperboloid respectively. We can still transform to Liouville normal form, if we first multiply the eigenoperator equation \eqref{eigenop} by a minus sign. Then we see that 
\be
  x=\int_{0}^R\frac{1}{\sqrt{|a_2(R')|}}\,dR'\,,
\ee
is the same transformation as before. The wave-function is now
\be
\label{psiminus}
  \psi(x)=|a_2|^{\frac{1}{4}}\!(R)\,w^{\frac{1}{2}}(R)\,v(R)\,,
\ee
whilst the Schr\"odinger equation now appears as
\be
\label{Schrominus}
  -\frac{d^2\psi(x)}{dx^2}+U(x)\,\psi(x)=-\lambda\, \psi(x)\,, 
\ee
\ie with $\lambda$ now being minus the energy. The potential $U$ is given by the same formula up to an overall sign, \ie
\be
  U(x)=-a_0+\frac{a_1^2}{4|a_2|}+\frac{a'_1}{2}-a'_2\Big(\frac{a_1}{2a_2}+\frac{3a'_2}{16a_2}\Big)+\frac{a''_2}{4}\,.
\ee

The power-law eigenoperators $v\sim |R|^{\frac\lambda2}$ are now associated with exponentially growing wave-functions, dominated by the $\omega$ dependence in \eqref{psiminus}. From Schr\"odinger's point of view, they are not acceptable solutions. On the other hand, the solutions $v\sim1/\omega(R)$ correspond to exponentially decaying $\psi(x)$ and thus bound-state solutions to \eqref{Schrominus}.

Since the large $R$ dependence of $a_1$ and $|a_2|$ is the same as before, we see that the analysis \eqref{aFG} -- \eqref{asympU} goes through unchanged and $U(x)$ has the same large $x$ dependence \eqref{largeU} as before. The WKB analysis therefore also goes through unchanged, except that the energies are now $-\lambda_n$. Therefore we see that we have at most a finite number of (marginally) irrelevant operators and an infinite tower of relevant operators, the scaling dimension of the conjugate couplings being
\be 
\label{thetaminus}
\theta_n = b\,(n\ln n)\left\{1+O\left(\frac{\ln\ln n}{\ln n}\right)\right\}\qquad\text{as}\qquad n\to\infty\,.
\ee

We recognise that these are $f(R)$-approximation analogues of the $\dd{k}n$ operators introduced in \cite{Morris:2018mhd} and studied extensively in refs. \cite{Kellett:2018loq,Morris:2018upm,Morris:2018axr,first,second} as elements of a new quantisation of quantum gravity. Indeed the $\dd{k}n$ operators are eigenoperators appearing in the functional RG when using  
a wrong-sign cutoff ($c_{\bar{h}}<0$), where it is needed because the conformal factor field, $\varphi$, has wrong-sign kinetic term. The $\dd{k}n$ span the space of perturbations that are square integrable under an exponentially growing SL measure, and thus are themselves exponentially decaying at large field. Finally they also form an infinite tower of relevant operators, the scaling dimensions being $\theta_n = 5+n$. 

Note however that the $\dd{k}n$ are eigenoperators about the Gaussian fixed point, where they are derived exactly, whereas the formula \eqref{thetaminus} applies to the spectrum of square-integrable eigenoperators about any point in the continuum of fixed points in this $f(R)$-approximation. Unlike the $\varphi$-versions, the eigenoperator equation has coefficients $a_i(R)$ with non-trivial field dependence. This is responsible for the $\ln n$ dependence in \eqref{thetaminus} while, as we noted in sec. \ref{sec:Liouville}, the $b$ dependence appearing in \eqref{thetaminus} is a symptom of the single-metric approximation.

\section{Summary and Conclusions}
\label{sec:conclusions}


We use the $f(R)$ model introduced in ref. \cite{Benedetti:2013jk} where already SL theory was applied to give a proof that, around any fixed point in such a model, there are a finite number of relevant couplings and an infinite number of irrelevant couplings $g_n$, these latter having scaling dimensions $\theta_n\to-\infty$ as $n\to\infty$. Note that the scaling dimensions are also proved to be real, in contrast to what is found typically in finite dimensional truncations.
In this paper we scrutinise both the explicit and implicit assumptions that go into this proof, and we combine SL techniques with asymptotic analysis at large $R$ \cite{Dietz:2012ic,Gonzalez_Martin_2017} to find out significantly more about the nature of these fixed points and their eigenoperator spectrum. 

Both of these methods can be developed while keeping the cutoff general, which must however be taken to be smooth. In  \eqref{eq:15} we keep general the $c_\phi$ (the overall size of the cutoff for each field component). As in ref. \cite{Benedetti:2013jk}, we set the endomorphism parameters $\alpha_2=\alpha_1=0$, but we keep $\alpha_0$ general apart from the constraint $1/3<\alpha_0<25/48$ required to ensure that all modes are integrated out in the limit $k\to0$. We take the same cutoff profile for all field components, since these are all closely tied to the metric either through changes of variables or via BRST invariance.
For most of the paper to be concrete we specialise to the exponential-style cutoff profile  \cite{Wetterich:1992} \eqref{Wetr}, but we keep its parameters $a>0$ and $b\ge1$ general.
In particular we are able to determine the asymptotic form of the SL weight $\omega(R)$ for these cases. It is a rapidly decaying exponential of an exponential \cf \eqref{omegaH} and \eqref{omegaS}  for the hyperboloid and sphere respectively.
We show that it is intimately involved in other asymptotic properties, chief amongst them being  the detailed form \eqref{largen} of the asymptotic behaviour of the $\theta_n$:
\be 
\label{largenagain}
\theta_n = -b\,(n\ln n)\left\{1+O\left(\frac{\ln\ln n}{\ln n}\right)\right\}\qquad\text{as}\qquad n\to\infty\,.
\ee
If computed exactly, these scaling dimensions should be universal. Thus it is gratifying to find that in this model approximation,  they are independent of all parameters except one within our general family of cutoffs. It is also encouraging to find that the $\theta_n$ have an almost linear dependence on $n$, since in this respect it is similar to the numerical evidence for near-Gaussian (but complex) dimensions found in ref. \cite{Falls:2018ylp} for $n\le 70$ in an adaptive optimised cutoff version of the $f(R)$ approximation. However the overall dependence on $b$ still amounts to strong residual cutoff dependence, precluding any more meaningful comparison. We saw in sec. \ref{sec:Liouville} that the blame for this lies squarely with the single metric approximation. In fact single field approximations are a known source of artefacts \cite{Bridle:2013sra}.

SL theory requires the RG eigenvalue equation \eqref{eigenop} to be second order in $R$ derivatives. This is achieved if and only if we use a non-adaptive cutoff profile. While that leads to the disadvantage of significantly more complicated flow equations compared to those using an adaptive optimised cutoff \cite{Litim:2001},
it does allow us also to ensure that the fixed point ODE has no fixed singularities. 

This is an advance on $f(R)$ approximations with adaptive cutoff, where such fixed singularities are endemic. While the fixed singularity at $R=0$ appears there for a clear physical reason \cite{Benedetti:2012,Dietz:2012ic}, the same is not true for those at $R\ne0$. These latter fixed singularities can be introduced or shifted to different places, depending on the model \cite{Demmel2015b,Ohta2016}, but it seems to be impossible to eliminate them entirely  \cite{Machado:2007,Codello:2008,Benedetti:2012,Demmel:2012ub,Demmel:2013myx,Demmel:2014hla,Demmel:2014fk,Demmel2015b,Ohta:2015efa,Ohta2016,Percacci:2016arh,Morris:2016spn,Falls:2016msz,Ohta:2017dsq,Benedetti:2013jk}. However, solutions depend sensitively on them, in particular determining whether fixed points exist as global solutions and if so whether they form a continuous set \cite{Dietz:2012ic,Gonzalez_Martin_2017}. 

On the other hand an adaptive cutoff profile has the advantage in that it adapts to the sign of the Hessian. In our case we have to fix the sign of the cutoff via $c_\phi$.
The Hessian is positive for nearly all field components, requiring $c_\phi>0$, as would anyway be expected for convergence of the functional integral. However the physical scalar component $\bar{h}$, \aka the conformal factor,  is an exception. If we are to describe the regime corresponding to perturbative quantisation of the Einstein-Hilbert term we need to choose $c_{\bar{h}}<0$ \cite{Reuter:1996,Dietz:2015owa,Dietz:2016gzg,Morris:2018mhd}. Otherwise we need to rely on $f_k(R)$ containing higher order terms \cite{Lauscher:2002sq} so that $f''_k(R)$ is positive, \cf \eqref{Th} and the discussion in sec. \ref{sec:fReqs} and at the beginning of sec. \ref{sec:flat}. We choose $c_{\bar{h}}>0$ for the body of the paper, following ref. \cite{Benedetti:2013jk}.

It turns out that on the sphere, we can find the leading asymptotic behaviour of the fixed point solution $f(R)$ in the large $R$ limit for completely general cutoff profile $r(z)$. The result, \eqref{asympGen}, is different from the assumed form in ref. \cite{Benedetti:2013jk}. In fact it is dominated by cutoff effects. For the exponential cutoff it takes the form \eqref{asympFPS}. As discussed in sec. \ref{sec:largeRfp}, this limit also ought to be universal, giving the physical equation of state. Here we saw that the blame lies squarely with the course-graining of constant scalar modes in the Jacobian of the change of variables to York decomposition. We saw that this had no effect on the $\theta_n$ formula \eqref{largenagain} however.

The asymptotic solution  contains one parameter, $A$, whereas for a second-order ODE we would expect a general solution to have two. By perturbing around this result we saw that to leading order the other parameter multiplies $\delta\!f(R)\sim1/\omega(R)$. Since this perturbation grows more rapidly than $f(R)$, it is not valid asymptotically and thus we see that asymptotically there is only a one-parameter set of fixed point solutions. As discussed in sec. \ref{sec:largeRfp} if we consider the flow equations as applying only to the sphere, we would then have line(s) of fixed points. This is one motivation for widening the domain of applicability of the flow equations.
As discussed in sec. \ref{sec:flat} nor would we be able to apply SL theory, the obstruction coming from the existence of an $R=0$ boundary (where the equations go over to those of flat space). This provides  another motivation. As a final motivation we appeal to the encouraging evidence found in polynomial approximations to $f(R)$ equations \cite{Codello:2007bd,Falls_2014,Falls:2018ylp,Falls:2017lst,Kluth:2020bdv}. These polynomials probe both signs of $R$. We saw at the end of sec. \ref{sec:flat} that if we wish to keep the same cutoff profile for all modes we cannot analytically continue our equations into $R<0$ however. Instead we match the solution into the equations on the hyperboloid, which also has the property that the equations go over to the flat space ones at its $R=0$ boundary. 

On the hyperboloid the leading asymptotic behaviour is cutoff independent as it should be, being $f(R)\sim AR^2$ (for a typically different $A$ compared to the sphere side). We also provided the leading corrections coming from cutoff terms \eqref{asympFPH}, as we did also on the sphere \eqref{asympFPS}. Again a perturbation to this solution takes the form $\delta\!f(R)\sim1/\omega(R)$ and is thus ruled out. Therefore the asymptotic behaviour as $R\to\pm\infty$ provides two constraints on a global solution for $f(R)$ leading to at most a discrete set of fixed points. This is of course what one would hope to see for asymptotic safety.\footnote{Note that had we introduced fixed singularities into the $f(R)$ equations we would then have found $f(R)$ to be overconstrained and have no global solutions.}

In sec. \ref{sec:largeReigen} we saw that the  situation is just as encouraging for the eigenoperators $v(R)$. Since in the eigenoperator equation \eqref{eigenop}, $a_2(R)$ vanishes asymptotically on both the sphere and the hyperboloid (for the explicit formulae see \eqref{asympa2S} and \eqref{asympa2H} respectively), the leading asymptotic behaviour for an eigenoperator is given by $v(R)\propto |R|^{\frac{\lambda}{2}}$, which is again universal, as it should be (if computed exactly). For any RG eigenvalue $\lambda$ the other solution grows rapidly with $|R|$, satisfying asymptotically $v(R)\sim1/\omega(R)$ (in agreement with $\delta\!f(R)$ which corresponds to a putative marginal operator). It is ruled out because it does not evolve multiplicatively under the RG. Since the ODE is linear second order, requiring $v(R)\propto |R|^{\frac{\lambda}{2}}$ overconstrains the equations and leads to quantisation of $\lambda$, again as one would hope to see. 

Furthermore these `power-law' eigenoperators are square-integrable under the SL weight, thus providing the missing justification for using SL analysis. From general SL theory, this is already enough to confirm that the eigenoperators $v_n(R)$ form a discrete spectrum and to show that the RG scaling dimensions $\lambda_n$, possibly finitely degenerate, have a finite minimum (thus there are a finite number of relevant directions) and form an infinite tower such that (ordering the eigenoperators so $\lambda_n$ are non-decreasing in $n$) the $\lambda_n\to\infty$ as $n\to\infty$. The $v_n(R)$ can be chosen to be orthonormal under the SL weight $\omega(R)$. In fact, the rest of the SL analysis in ref. \cite{Morris:1996xq} can then be straightforwardly taken over to show that arbitrary bare perturbations $\delta\!f_{k_0}(R)$ (at some UV scale $k=k_0$) will evolve into the space of interactions that can be expanded over the $v_n(R)$ such that the series converges in the square-integrable sense. The map to Liouville normal form, done in sec. \ref{sec:Liouville} and ref. \cite{Benedetti:2013jk}, allows us to take this further by computing the large distance behaviour \eqref{largeU} of its potential, and from there, by a standard application of WKB analysis, to derive the asymptotic form \eqref{largenagain} of the $\theta_n=4-\lambda_n$ as quoted above.

All this is predicated on there actually being a global solution to the fixed point equation \eqref{fixed} however. We have searched numerically for such a solution in the case $a=b=1$, $\alpha_0=1/2$ (recall from sec. \ref{sec:hyperboloid} that it has to lie between $1/3$ and $25/48$) and all the $c_\phi=1$.  We found global solutions on the sphere that asymptote to \eqref{asympFPS} for a small region around $A=-0.01$,  starting at $R=10$ and integrating down to the flat space fixed point equation from \eqref{flatflow}, but we have not been able to find global solutions on the hyperboloid. These are challenging integro-differential (on the sphere-side sum-differential) equations so it is likely that more numerical work is required. This includes exploring other choices of parameters.  In fact our solutions on the sphere matched the asymptotic solution \eqref{asympFPS} at  $R=10$, only by choosing to match $f'(R)$ and $f''(R)$ and then computing $f(R)$ from the fixed point equation (rather than the more obvious route of  setting $f(R)$ and $f'(R)$ from the asymptotic formula). This indicates that the asymptotic series has not been taken  quite far enough for these $R$ values. On the hyperboloid, the asymptotic corrections in \eqref{asympFPH} fall only slowly, so would surely have to go much further to provide a similarly accurate starting point. In fact it would be beneficial to explore simpler equations, if these can be found. An attractive starting point would be to use non-adaptive cutoff together with the exponential parametrisation explored in ref.  \cite{Ohta:2015efa}. Note that if lines of fixed points can be found on both sphere and hyperboloid, there would still have to be a matching point where these $f(R)$ agree to second order in their Taylor expansion \eqref{Taylor} about $R=0$, in order to have found a globally defined fixed point.

Finally in sec. \ref{sec:negative} we saw that the situation is dramatically different if we choose instead the wrong sign cutoff for the conformal mode: $c_{\bar{h}}<0$. Perturbing around the asymptotic fixed point solution we still find $\delta\!f(R)\sim1/\omega(R)$, but the dependence of the SL measure on $c_{\bar{h}}$ is such that $\omega(R)$ is now a rapidly growing exponential of an exponential. This means that the perturbation $\delta\!f(R)$ remains valid asymptotically, and thus the asymptotic solutions have two parameters. They no longer restrict the dimension of the solution space, so fixed points form two-dimensional continuous sets. The alternative asymptotic behaviour for the eigenoperators is also still $v(R)\sim 1/\omega(R)$ but these do now evolve multiplicatively under the RG and thus are also valid solutions. Therefore we have a non-quantised continuous spectrum of RG eigenvalues $\lambda$. It is clear that this is mirroring  effects previously found \cite{Dietz:2012ic,Gonzalez_Martin_2017} in adaptive cutoff $f(R)$ approximations  \cite{Benedetti:2012,Demmel2015b}, and found \cite{Dietz:2016gzg} in a background-independent version of the so-called conformally reduced gravity \cite{Dietz:2015owa} where only the conformal factor field is kept. Clearly therefore the culprit for this degeneration is the wrong sign cutoff (which is necessary however if we work with wrong sign kinetic term). In this case, by choosing to keep only interactions square integrable under the SL measure, the eigenoperator spectrum is again quantised, with $v(R)\sim 1/\omega(R)$ for large $R$. These form a tower of operators, only finitely many of which are irrelevant, and infinitely many are relevant with dimensions given by minus the $\theta_n$ in \eqref{largenagain}. These are the $f(R)$-approximation analogues of the $\dd{k}n$ operators pursued in \cite{Morris:2018mhd,Kellett:2018loq,Morris:2018upm,Morris:2018axr,first,second} as an alternative quantisation of quantum gravity.

\vskip2cm

\section*{Acknowledgments}

AM and DS acknowledge support via STFC PhD studentships. TRM acknowledges support from STFC through Consolidated Grant ST/T000775/1.

\newpage

\bibliographystyle{hunsrt}
\bibliography{references}
\end{document}